\documentclass[10pt,prl,superscriptaddress,showpacs]{revtex4}
\usepackage{amsmath}
\usepackage[T1]{fontenc}
\usepackage{amsfonts}
\usepackage{amssymb}
\newcommand{\oq}{\frac{1}{4}}
\newcommand{\oh}{\frac{1}{2}}
\newcommand{\tq}{\frac{3}{4}}
\begin{document}
\title{Explicit form of correlation function three-settings tight Bell inequalities for three qubits}
\author{Marcin Wie\'sniak}\affiliation{
Institute of Theoretical Physics and Astrophysics, University
of Gda\'nsk, PL-80-952 Gda\'nsk}
\author{Piotr Badzi\k{a}g}\affiliation{Alba Nova Fysikum, University of Stockholm, S–106 91, Sweden}\affiliation{
Institute of Theoretical Physics and Astrophysics, University
of Gda\'nsk, PL-80-952 Gda\'nsk}
\author{Marek \.Zukowski}\affiliation{
Institute of Theoretical Physics and Astrophysics, University
of Gda\'nsk, PL-80-952 Gda\'nsk}\affiliation{Alba Nova Fysikum, University of Stockholm, S–106 91, Sweden}

\begin{abstract}
We present a method to derive explicit forms of tight correlation function Bell
inequalities for three systems and dichotomic observables, which involve three settings for each observer.
We also give sufficient and necessary conditions for quantum predictions to satisfy the new inequalities.
\end{abstract}

\maketitle

\section{Introduction}
The problem of the possibility of a local realistic interpretation 
of quantum mechanics was first addressed in the discussion between
Einstein, Podolsky, Rosen \cite{epr1935} and Bohr \cite{bohr1935}. 
The final answer was provided in 1964 by Bell \cite{bell}.
The paper of Bell contains an inequality for local realistic (LR) correlation functions, which
can be violated only by non-classical, entangled states. The correlation function is defined as a mean value of a product of local observables. For example, a two-qubit correlation function shall be given by $E_{ij}=\langle A_iB_j\rangle$, with $A_i$ being an $i$th observable for the first observer (Alice) and $B_j$ is $j$th for the second (Bob). The observables are dichotomic, with eigenvalues $\pm 1$. In the work of Clauser,
Horne, Shimony, and Holt (CHSH) \cite{CHSH} a new inequality was derived, which, in contradistinction to the original Bell's expression, can be applied to real physical processes.\footnote{The original Bell inequality cannot be derived without assuming perfect (anti-)correlations. Since every experiment involves some imperfections, experimental data never agree with this assumption. } Again, it was formulated for LR correlation functions. Such formulations of the Bell inequalities are now the most
popular forms of it. The discovery of the Greenberger-Horne-Zeilinger correlations triggered generalizations of  Bell inequalities to more subsystems \cite{Mermin}. The early efforts in this direction were restricted to the scenarios involving two experimental settings per observer only. The full set of such inequalities was finally given in Refs. \cite{www,zb}. More recently, however, maps between general Bell inequalities for correlation functions (even yet unknown ones) and quantum cryptography \cite{auz}, as well as quantum communication complexity problems \cite{bzpz}, have been found. Thus, there is an additional motivation to find  new {\em tight} Bell inequalities of this kind, with the obvious direction of generalization toward more allowed settings for each observer. Allowing more settings leads e.g. to more complicated communication complexity problems, and from the fundamental point of view, must further restrict the realm of classically admissible correlations. 

By tight Bell inequalities one understands these, which define walls of a
polytope in a certain
statistical hyperspace, which in turn defines the full set of possible local-realistic predictions for the values of correlation functions for the given set of Bell-type experiments. The vertices of the polytope are 
deterministic predictions, that is, are given by extremal correlation functions for which the product of the  values for the each pair of measurement directions are always the same. Thus vertices have components equal to $\pm1$. 
The polytope was initially proposed for the probability hyperspace \cite{pitowski}. Various aspects of the problem were subsequently discussed by a number of authors, for example in \cite{tsirelson,pitowskisvozil,sliwa}. More recently,
a method to derive the general structure of coefficients of tight Bell inequalities for correlation functions, which involve three, or more, settings
available to each observer, was presented in \cite{ZUKOWSKI}. 
Here, basing on the results of \cite{ZUKOWSKI} we develop this method with the aim of getting the {\em explicit} form of the 
inequalities for three observers.
This is the simplest case for which tight three-setting Bell inequalities are thus far  unknown. 

\section{Three qubit inequalities}
Let us consider a situation, in which each of the three observers can choose
between three measurements. For three systems (observed by
Alice, Bob and Carol independently) and dichotomic observables, the tight Bell inequalities have a general form of \cite{ZUKOWSKI}
\begin{eqnarray}
\label{k}
-1\leq&\frac{1}{2^9}\sum_{a_i,b_j,c_k=\pm 1}S(a_0,a_1,a_2;b_0,b_1,b_2;c_0,c_1,c_2)&\nonumber\\
&\times\left(\sum_{i,j,k=0}^2a_ib_jc_kE_{ijk}\right)&\leq 1.\end{eqnarray}
where
$E_{ijk}$ is the value of the local realistic correlation function (i.e. the average of the product of the local results, which in turn are assumed to be $\pm1$) for settings $i,j,k$ of $A$, $B$, and $C$. $S(...)$ is a sign function (values $\pm1$), of the nine parameters, $a_0,...,c_2$. 
An admissible sign function has the form of \cite{ZUKOWSKI}
\begin{eqnarray}
&S(a_0,a_1,a_2;b_0,b_1,b_2;c_0,c_1,c_2)& \nonumber\\
&=\sum_{i,j,k=0}^2g_{ijk}a_ib_jc_k=\pm 1.
\label{S}
\end{eqnarray}
This relation can be inverted to give 
\begin{equation}
\label{ineqa}
g_{ijk}=\frac{1}{2^9}\sum_{a_i,b_j,c_k=\pm 1}S(a_0,a_1,a_2;b_0,b_1,b_2;c_0,c_1,c_2)a_ib_jc_k.
\end{equation}
This allows  rewriting (\ref{k}) as:
\begin{equation}
\left|\sum_{i,j,k=0}^2g_{ijk}E_{ijk}\right|\leq 1,
\end{equation}
where $E_{ijk}=\langle A_iB_jC_k\rangle$. 

Note also that the form of the sign functions, as in equation (\ref{S}), restricts the admissible class which generates the inequalities. The elementary trait of the class is that the admissible sign functions do not depend upon product of indices pertaining to one observer, e.g. $a_1a_2$, etc.  Note also, that one can reduce the number of considered inequalities to those involving sign functions with $a_0=1$, and in such a case the normalization factor changes from $\frac{1}{2^9}$ to $\frac{1}{2^6}$. This convention was used in ref. \cite{ZUKOWSKI}. However, it turns out that the more symmetric convention is handier for pinpointing the exact form of the coefficients, and thus we shall adopt it here.

\section{Properties of the coefficients in the inequalities}

To derive explicit Bell expressions it is convenient to define the following quantities, which we will further on call deltas. The first order deltas:
\begin{equation}
\Delta_x=\frac{1}{2}(S(x=1)-S(x=-1)),
\end{equation}
the second order deltas:
\begin{equation}
\Delta_{xy}=\frac{1}{2}(\Delta_x(y=1)-\Delta_x(y=-1)),
\end{equation}
and the third order deltas:
\begin{equation}
\Delta_{xyz}=\frac{1}{2}(\Delta_{xy}(z=1)-\Delta_{xy}(z=-1)).
\end{equation}
Each of the indices $x,y,z$ is related to a different party, e. g. $x=a_0,a_1,a_2; y=b_0,b_1,b_2; z=c_0,c_1,c_2$.

The third order deltas are equal to the coefficients $g_{jkl}$.
We can create the second order deltas from them in the following way:
\begin{widetext}
\begin{equation}
\Delta_{a_0b_0}(c_0,c_1,c_2)=c_0\Delta_{a_0b_0c_0}+c_1\Delta_{a_0b_0c_1}+c_2\Delta_{a_0b_0c_2}=c_0g_{000}+c_1g_{001}+c_1g_{002},
\end{equation}
and likewise for the remaining $\Delta_{a_ib_j}$'s. Likewise, 
\begin{eqnarray}
&\Delta_{a_0}(b_0,b_1,b_2;c_0,c_1,c_2)=b_0\Delta_{a_0b_0}(c_0,c_1,c_2)+b_1\Delta_{a_0b_1}(c_0,c_1,c_2)+b_2\Delta_{a_0b_2}(c_0,c_1,c_2),&\nonumber \\
\end{eqnarray}
and
\begin{eqnarray}
&S(a_0,a_1,a_2;b_0,b_1,b_2;c_0,c_1,c_2)&\nonumber \\
&=a_0\Delta_{a_0}(b_0,b_1,b_2;c_0,c_1,c_2)+a_1\Delta_{a_1}(b_0,b_1,b_2;c_0,c_1,c_2)+a_2\Delta_{a_2}(b_0,b_1,b_2;c_0,c_1,c_2)&,
\end{eqnarray}
\end{widetext}
etc..

\subsection{Properties of the deltas}
As $S=\pm 1$, each of the first order deltas $\Delta_{a_i}$ can be equal
to $\pm 1$ ($S$ changes with $a_i$) or 0 ($S$ does not change with $a_i$).
From the possible values of the first order deltas we can deduce, that a
second order delta can only be $0$ (no change), $\pm \frac{1}{2}$
(a change of $\Delta_{a_i}$ with $a_i$ from 0 to $\pm 1$, or vice versa), or
$\pm 1$ (a change between 1 and $-1$). By the same argument the coefficients $g_{ijk}$ may assume values $0,\pm\oq, \pm\oh, \pm\tq$ or $\pm 1$.


Finally, one can easily see that the coefficients $g_{ijk}$  satisfy the following relation 
$$\left|\sum_{j,k,l=0}^2g_{jkl}\right|=\sum_{j,k,l=0}^2g^2_{jkl}=1.$$
The first condition is due to a fact that the inequalities we are
searching for are saturated for all deterministic LR models, the
second is related to Parseval's theorem: for every sign
function of nine dichotomic variables we have $\sum_{a_i,b_j,c_k=\pm 1}S^2=512$.

The last inequality immediately implies that when one of coefficients $g_{ijk}$ has modulus 1, then all the others must be 0. In such a case, we get a family of trivial
inequalities $|E_{ijk}|\leq 1$.

\subsection{Construction of  sign functions}
Each second order delta can
contain one, two, or three coefficients $g_{ijk}$, and, similarly,
the first order deltas contain up to three second order deltas.
Not every pair or triplet of the second order deltas can be arranged into
higher-order deltas, however. The following table lists the families
of $\Delta_{a_ib_j}$'s, and what other $\Delta_{a_ib_j}$'s they can
go with to create a valid $\Delta_{a_i}$ (we avoid repetitions in
the table):
\begin{widetext}
\begin{tabular}{|c|c|}
\hline $\Delta_{a_ib_j}$&goes with:\\ \hline $\oq((-1)^m3c_x+(-1)^nc_y)$&$\pm\oq((-1)^mc_x-(-1)^nc_y)$\\ \hline
$\oq(2(-1)^mc_x+(-1)^nc_y+(1)^oc_z)$&$\pm\oq((-1)^mc_x-(-1)^nc_y)$ or $\pm\oq(2(-1)^mc_x-(-1)^nc_y-(-1)^oc_z)$ or\\& $\pm\oq((-1)^mc_x-(-1)^nc_y)$ and $\pm\oq((-1)^mc_x-(-1)^oc_z)$\\ \hline
$\oh((-1)^mc_x+(-1)^nc_y)$& alone or with $\oh((-1)^mc_x-(-1)^nc_y)$ or\\& two $\oq((-1)^mc_x-(-1)^nc_y)$'s\\ \hline $\oh(-1)^mc_x$& any one of $\pm\oh c_x,\pm\oh
 c_y,\pm\oh c_z$ or $\pm\oq((-1)^mc_x+(-1)^nc_y)$ and \\& $\pm\oq((-1)^mc_x-(-1)^nc_y)$ or $\pm\oq((-1)^nc_y+(-1)^oc_z)$ and $\pm\oq((-1)^nc_y-(-1)^oc_z)$\\ \hline
$\oq((-1)^mc_x+(-1)^nc_y)$&$\pm\oq((-1)^mc_x+(-1)^nc_y)$ or $\pm\oq((-1)^nc_y-(-1)^oc_z)$ and $\pm\oq((-1)^mc_x+(-1)^oc_z)$\\ \hline
\end{tabular}

\vspace{3mm}

\end{widetext}
In the table $(x,y,z)$ stand for a permutations of $(0,1,2)$, while $m,n,o\in \{0,1\}$.

Thus any first order delta belongs (after some transformations, i.
e. permutations of observables, or sign flips) to one of the families
listed below:
\begin{widetext}
\begin{tabular}{|c|c|c|c|c|c|}
\hline
 &$\Delta_{a_i}$&$|\Delta_{a_i}|^2$&  &$\Delta_{a_i}$&$|\Delta_{a_i}|^2$\\
\hline $\Delta_0$&$\oh (b_0(c_0+c_1)+b_1(c_0-c_1))$&$\frac{16}{16}$&$\Delta_I$&$\oq(b_0(-3c_0+c_1)+b_1(c_0+c_1))$&$\frac{12}{16}$\\ \hline
$\Delta_{II}$&$\oh b_0(c_0+c_1)$&$\frac{8}{16}$&$\Delta_{III}$&$\oh(b_0+b_1)c_0$&$\frac{8}{16}$\\ \hline
$\Delta_{IV}$&$\oh(b_0c_0+b_1c_1)$&$\frac{8}{16}$&$\Delta_{V}$&$\oq(b_0(2c_0+c_1+c_2)+b_1(c_1-c_2))$&$\frac{8}{16}$\\ \hline
$\Delta_{VI}$&$\oq(b_0+b_1)(c_0+c_1)$&$\frac{4}{16}$&$\Delta_{VII}$&$\oq(b_0(c_1-c_2)+b_1(c_0-c_1)+b_2(c_0-c_2))$&$\frac{6}{16}$\\
\hline
$\Delta_{VIII}$&$\oq(b_0(2c_0+c_1+c_2)+b_1(c_0-c_1)+b_2(c_0-c_2))$&$\frac{10}{16}$&$\Delta_{IX}$&$\oq(b_0(2c_0+c_1+c_2)+b_1(2c_0-c_1-c_2))$&$\frac{12}{16}$\\
\hline
$\Delta_X$&$\oq(b_0(2c_0+2c_1)+(b_1+b_2)(c_0-c_1))$&$\frac{12}{16}$&$\Delta_{XI}$&$\oq(2b_0+b_1(c_0+c_1)+b_2(c_0-c_1)$
&$\frac{8}{16}$\\ \hline $\Delta_{XII}$&$\frac{1}{4}(2b_0c_0+b_1(c_1+c_2)+b_2(c_1-c_2))$&$\frac{8}{16}$& & & \\
\hline
\end{tabular}
\vspace{3mm}
\end{widetext}
The norm $|\Delta_{a_i}|^2$ is the sum of squares of those $g_{ijk}$ coefficients, which enter a given delta. As it has been already
mentioned before, in an admissible sign function these norms add up to 1 \cite{endnote}.
However, this is only a necessary condition to build an inequality. The necessary and sufficient condition is that $\sum_{i=0}^2|\Delta_{a_i}|^2=1$ and $\Delta_{a_i}\Delta_{a_{i'}}=0$ for $i\neq i'$. Only a few pairs or
triplets of these deltas lead us to a proper Bell expression.

\section{Two qubits}
As an illustration of the method we will consider the two qubit case first. 

For two systems (observed by
Alice and Bob), the tight Bell inequalities have a general form of \cite{ZUKOWSKI}
\begin{eqnarray}
\label{k1}
-1\leq&\frac{1}{2^6}\sum_{a_i,b_j=\pm 1}S(a_0,a_1,a_2;b_0,b_1,b_2)\left(\sum_{i,j=0}^2a_ib_jE_{ij}\right)&\leq 1.\nonumber\\
\end{eqnarray}
where
$E_{ij}=\langle A_i B_j\rangle$ is the value of the correlation function for the $i$th and $j$th setting of Alice and Bob, respectively. $S(...)$ is the sign function of $a_0,a_1,a_2,b_0,b_1,b_2$. The coefficients $g_{ij}$'s of this function are related to $S$ by \cite{ZUKOWSKI}:
\begin{eqnarray}
&S(a_0,a_1,a_2;b_0,b_1,b_2)=\sum_{i,j=0}^2g_{ij}a_ib_j=\pm 1.& \nonumber \\
\label{S1}
\end{eqnarray}
Thus
\begin{equation}
\label{ineqb}
g_{ij}=\frac{1}{2^6}\sum_{a_i,b_j=\pm 1}S(a_0,a_1,a_2;b_0,b_1,b_2)a_ib_j.
\end{equation}
This allows to rewrite (\ref{k1}) as
\begin{equation}
\left|\sum_{i,j=0}^2g_{ij}E_{ij}\right|\leq 1.
\end{equation}

It is enough to define first and second order deltas. The second order deltas are equivalent to $g_{ij}$'s.
As we have mentioned, the first order deltas can only take values of $0$ or $\pm 1$ and $g_{ij}=0,\pm\oh,$ or $\pm 1$. If some $\Delta_{a_i}$ is always equal to $\pm 1$, the sign function is factorisable, and generates a trivial bound:
\begin{equation}
\label{q21}
\left|E_{ij}\right|\leq 1.
\end{equation}
If some $\Delta_{a_i}$ takes all three values ($\pm 1$ and 0), it can be locally transformed to $\oh(b_0+b_1)$. The other $\Delta_{a_i}$ vanishes when the first one does not, and vice versa, so it can be taken as $\oh(b_0-b_1)$. Thus we reproduce the CHSH inequality \cite{CHSH}:
\begin{equation}
\label{q22}
\frac{1}{2}\left|E_{00}+E_{01}+E_{10}-E_{11}\right|\leq 1.
\end{equation}
No other inequalities of this kind for two qubits are allowed, because we have utilized all possible values of $g_{ij}$'s and all possible forms of $\Delta_{a_i}$'s. 

As we see the set of Bell inequalities for correlation functions, for two partner and three settings on each side, boils down to the good old CHSH inequalities, for all possible combinations of pairs of settings. This result corroborates with the earlier finding of Garg \cite{GARG}. 

\section{Three-qubit inequalities}
Since the method is universal, it produces both the well-known standard Bell inequalities \cite{www,zb}, which are shown in the first subsection below, and the new ones, which are the subject of the second subsection.

\subsection{$2\times 2\times 2$ inequalities}
\begin{itemize}
\item
Using $\Delta_I$ and $\Delta_{IV}$ we can obtain the following sign function:
\begin{eqnarray}
S=&\oq(a_0(b_0(-3c_0+c_1)+b_1(c_0+c_1))+a_1(b_0+b_1)(c_0+c_1)),
\end{eqnarray}
and by putting it into (\ref{ineqa}), we obtain:
\begin{eqnarray}
\oq|-3E_{000}+E_{001}+E_{010}+E_{011}+E_{100}+E_{101}+E_{110}+E_{111}|&\leq 1.
\end{eqnarray}
\item
Taking two $\Delta_{II}$'s or two $\Delta_{III}$'s we get 
\begin{eqnarray}
S=&\oh(a_0b_0(c_0+c_1)+a_1b_0(c_0-c_1)).
\end{eqnarray}
This leads to the inequality, which is a trivial extension of the CHSH one:
\begin{eqnarray}
\oh|E_{000}+E_{001}+E_{100}-E_{101}|\leq 1,
\end{eqnarray}
With two $\Delta_{II}$'s or two $\Delta_{III}$'s one can also get
\begin{eqnarray}
S=&\oh(a_0(b_0+b_1)c_0+a_1(b_0-b_1)c_0),
\end{eqnarray}
this gives
\begin{eqnarray}
\oh|E_{000}+E_{010}+E_{100}-E_{110}|\leq 1.
\end{eqnarray}
\item
and
\begin{eqnarray}
S&=&\oh(a_0b_0(c_0+c_1)+a_1b_1(c_0-c_1)),
\end{eqnarray}
this in turn implies
\begin{eqnarray}
\label{ineq0}
\oh|E_{000}+E_{001}+E_{110}-E_{111}|\leq 1.
\end{eqnarray}
\item
Finally, a combination of $\Delta_{II}$ and $\Delta_{III}$ leads to
\begin{eqnarray}
S&=&\oh(a_0(b_0c_0+b_1c_1)+a_1(b_1c_0-b_0c_1)).
\end{eqnarray}
It results in a Mermin-Ardehali-Belinskii-Klyshko inequality \cite{belinskiklyshko}:
\begin{eqnarray}
\oh|E_{000}+E_{011}+E_{101}-E_{110}|\leq 1.
\end{eqnarray}

\end{itemize}
This closes the group of possible inequalities which involve only two settings (out of the three ones allowed in the considered scenario). 
\subsection{Inequalities with up to three measurement settings per observer}
However, our method can also lead to thus far unknown inequalities. The most important is one in which all three observers perform three alternative measurements. The sign function is built of $\Delta_V$ and two $\Delta_{VII}$'s:
\begin{eqnarray}
\label{signbadziag}
&S=\frac{1}{4}(a_0(b_1(c_1-c_2)+b_2(c_1-c_2))+a_1(b_0(c_1+c_2)+b_1(c_0+c_1)+b_2(-c_0+c_2))&\nonumber\\
 &+a_2(b_0(-c_1-c_2)+b_1(c_0+c_2)+b_2(-c_0+c_1))).&
\end{eqnarray}
It represents the following inequality:
\begin{eqnarray}
\label{badziagineq}
&\frac{1}{4}|E_{011}-E_{012}+E_{021}-E_{022}+E_{101}+E_{102}+E_{110}+E_{111}&\nonumber\\
&-E_{120}+E_{122}-E_{201}-E_{202}+E_{210}+E_{211}-E_{220}+E_{221}|&\leq 1.
\end{eqnarray}

From this inequality follow two other, less general conditions. Putting $a_0=a_2$ we get
\begin{eqnarray}
\label{sign1}
&S=\frac{1}{4}(-a_0(b_0(c_1+c_2)-b_1(c_0+c_1)+b_2(c_0-2c_1+c2)&\nonumber\\
&+a_1(b_0(c_1+c_2)+b_1(c_0+c_1)+b_2(-c_0+c_2))),&
\end{eqnarray}
this represents
\begin{eqnarray}
\label{ineq1}
&\frac{1}{4}|-E_{001}-E_{002}+E_{010}+E_{011}-E_{020}+2E_{021}&\nonumber\\
&+E_{022}+E_{101}+E_{102}+E_{110}+E_{111}-E_{120}+E_{222}|\leq 1.&
\end{eqnarray}
If we rather put $a_2=a_1$ we have
\begin{eqnarray}
\label{sign2}
&S=\frac{1}{4}(a_0(b_1(c_1-c_2)+b_2(c_1-c_2)&\nonumber\\
&+a_1(b_1(2c_0+c_1+c_2)+b_2(2c_0-c_1-c_2))),&
\end{eqnarray} 
this gives 
\begin{eqnarray}
\label{ineq2}
&\frac{1}{4}|E_{011}-E_{012}+E_{021}-E_{022}+2E_{110}&\nonumber\\
&+E_{111}+E_{112}+2E_{120}-E_{121}-E_{122}|&\leq 1.
\end{eqnarray}
Our method leads to yet another Bell inequality. Two $\Delta_{XII}$'s can be arranged to
\begin{eqnarray}
\label{sign3}
&S=\frac{1}{4}(a_0(2b_0c_0+b_1(c_1+c_2)+b_2(c_1-c_2))&\nonumber\\
 &+a_1(2b_0c_0-b_1(c_1+c_2)-b_2(c_1-c_2))),&
\end{eqnarray} 
what represents
\begin{eqnarray}
\label{ineq3}
&\frac{1}{4}|2E_{000}+E_{011}+E_{012}+E_{021}-E_{022}&\nonumber\\
&+2E_{100}-E_{111}-E_{112}-E_{121}+E_{122}|&\leq 1.
\end{eqnarray}
Interestingly, inequalities (\ref{ineq1}), (\ref{ineq2}), and (\ref{ineq3}) are special cases of the one given in \cite{wuzong,PLZ},
\begin{eqnarray}
\label{ineq244}
&\frac{1}{4}|E_{000}+E_{001}+E_{010}-E_{011}&\nonumber\\
&+E_{022}+E_{023}+E_{032}-E_{033}&\nonumber\\
&+E_{100}+E_{101}+E_{110}-E_{111}&\nonumber\\
&-E_{122}-E_{123}-E_{132}+E_{133}&\leq 1.
\end{eqnarray}
Inequality (\ref{ineq3}) is obtained by putting $\hat{C}_0=\hat{C}_1$. Then we can get (\ref{ineq2}) by $\hat{B}_0=\hat{B}_2$. Finally, choosing $\hat{B}_0=\hat{B}_2$ and $\hat{C}_0=\hat{C}_2$ from (\ref{ineq244}) we reach (\ref{ineq1}) Further such simplifications lead to the standard $2\times 2\times 2$ inequalities.
\section{Conditions on state to satisfy the new inequalities}
In this Section we will derive sufficient conditions which a state must satisfy in order to never violate the Bell inequality. Such conditions are not dependent on particular measurement settings, but refer only to the properties of the state.

As inequalities (\ref{ineq1}) and (\ref{ineq2}) are special cases of (\ref{badziagineq}), we focus only on the conditions for (\ref{badziagineq}) and (\ref{ineq3}). 

Let us  recall that any observable on a qubit with spectrum $\{-1,1\}$ can be written as $\hat{X}_i=\vec{X}_i\cdot\vec{\sigma}\!^{X}$ ($X=A,B,C$; $i=0,1,2$, the upper index of the Pauli matrix vector enumerates the qubit and all the vectors are normalized). In terms of the correlation tensor (elements of which are given by $T_{ijk}=Tr\rho(\sigma_{i}^A\otimes\sigma_{j}^B\otimes\sigma_{k}^C)$) inequality (\ref{badziagineq}) can be written as
\begin{eqnarray}
\label{badziag2}
&\frac{1}{4}\left|\hat{T}\odot\left(\vec{A}_0\otimes(\vec{B}_1+\vec{B_2})(\vec{C}_1-\vec{C}_2)\right.\right. &\nonumber\\
&+(\vec{A}_1-\vec{A}_2)\otimes\vec{B}_0\otimes(\vec{C}_1+\vec{C}_2)&\nonumber\\
&+(\vec{A}_1+\vec{A}_2)\otimes(\vec{B}_1+\vec{B_2})\otimes\vec{C}_0&\nonumber\\
&+\frac{1}{2}(\vec{A}_1+\vec{A}_2)\otimes(\vec{B}_1+\vec{B}_2)\otimes(\vec{C}_1+\vec{C}_2)&\nonumber\\
&\left.\left.+\frac{1}{2}(\vec{A}_1-\vec{A}_2)\otimes(\vec{B}_1-\vec{B}_2)\otimes(\vec{C}_1-\vec{C}_2)\right)\right| &\leq 1,
\end{eqnarray}
where the tensor scalar product is taken as $\hat{A}\odot\hat{B}=\sum_{i,j,k=1}^3A_{ijk}B_{ijk}$ Note that the tensor in (\ref{badziag2}) constructed from the vectors defining the measurements contains five mutually orthogonal (in the sense of $\odot$) terms. 

We now choose local coordinate systems, in which
\begin{eqnarray}
\vec{A}_1+\vec{A}_2=2\cos\alpha\vec{e}_1,\\
\vec{A}_1-\vec{A}_2=2\sin\alpha\vec{e}_2,\\
\vec{B}_1+\vec{B}_2=2\cos\beta\vec{e}_1,\\
\vec{B}_1-\vec{B}_2=2\sin\beta\vec{e}_2,\\
\vec{C}_1+\vec{C}_2=2\cos\gamma\vec{e}_1,\\
\vec{C}_1-\vec{C}_2=2\sin\gamma\vec{e}_2,
\end{eqnarray}
where $\vec{e}_1=(1,0,0)$, $\vec{e}_2=(0,1,0)$, and $\vec{e}_3=(0,0,1)$.
Let us moreover introduce a short-hand notation $T_{Aij}=\hat T\odot\vec{A}_0\otimes\vec{e}_i\otimes\vec{e}_j$, $T_{iBj}=\hat T\odot\vec{e}_i\otimes\vec{B}_0\otimes\vec{e}_j$, and $T_{ijC}=\hat T\odot\vec{e}_i\otimes\vec{e}_j\otimes\vec{C}_0$. This allows to rewrite (\ref{badziag2}) in a  form of 
scalar product of two real vectors:
\begin{eqnarray}
\label{badziag3}
&|(T_{A12},T_{2B1},T_{12C},T_{111},T_{222})&\nonumber\\
&\cdot(\cos\beta\sin\gamma,\sin\alpha\cos\gamma,\cos\alpha\sin\beta,\cos\alpha\cos\beta\cos\gamma,\sin\alpha\sin\beta\sin\gamma)|\leq 1.&
\end{eqnarray} 
This is the necessary and sufficient condition for the given state to satisfy the inequality. That is, if the maximum of the left hand side for all possible local coordinate systems,  vectors $\vec{A}_0$,
$\vec{B}_0$ and $\vec{C}_0$, and angles $\alpha$, $\beta$ and $\gamma$ is less or equal to $1$, then the predictions for given state always satisfy the inequality. 

If one employs the Cauchy inequality, $|\vec{v}_1\cdot\vec{v}_2|^2\leq|\vec{v}_1|^2|\vec{v}_2|^2$, one can 
formulate a concise {\em sufficient} condition on a state to satisfy (\ref{badziagineq}). The state cannot violate the inequality if
\begin{eqnarray}
\label{badziagcond1}
T_{111}^2+T_{222}^2+T_{A12}^2+T_{2B1}^2+T_{12C}^2\leq 1
\end{eqnarray}
holds in all local bases and for all $\vec{A}_0,\vec{B}_0,\vec{C}_0$. Finally, we utilize the fact that $\max_{|\vec{A}_0|=1}T_{A12}^2=\sum_{i=1}^3T_{i12}^2$ (this is because the optimal vector is  $\vec{A}_0^{opt}=(T_{112},T_{212},T_{312})/\sqrt{T_{112}^2+T_{212}^2+T_{312}^2}$), etc., and obtain the following form:
\begin{equation}
\label{badziagcond2}
T_{111}^2+T_{222}^2+\sum_{i=1}^3(T_{i12}^2+T_{2i1}^2+T_{12i}^2)\leq 1.
\end{equation}

Let us derive a similar condition for (\ref{ineq3}). In the fashion of (\ref{badziag3}) the inequality reads
\begin{eqnarray}
\label{ineq32}
&|\hat{T}\odot(\cos\alpha\vec{e}_1\otimes\vec{B}_0\otimes\vec{C}_0&\nonumber\\
&+\sin\alpha\vec{e}_2\otimes(\cos\beta\cos\gamma\vec{e}_1\otimes\vec{e}_1+\cos\beta\sin\gamma\vec{e}_1\otimes\vec{e}_2&\nonumber\\
&+\sin\beta\cos\gamma\vec{e}_2\otimes\vec{e}_1-\sin\beta\sin\gamma\vec{e}_2\otimes\vec{e}_2)|&\leq 1
\end{eqnarray}
(we have renumbered observables of Alice). Now, repeating the above argument, we get a necessary and sufficient condition on a state to satisfy (\ref{ineq3}). Namely in all local coordinate systems and for all $\vec{B}_0$ and $\vec{C}_0$ we shall have
\begin{equation}
T^2_{1BC}+\sum_{i,j=1}^2T_{2ij}^2\leq 1,
\end{equation}
or using the trick presented earlier
\begin{equation}
\sum_{k=1}^3T^2_{1kC}+\sum_{i,j=1}^2T_{2ij}^2\leq 1.
\end{equation}
\section{Conclusions}
A  method to derive explicit Bell inequalities with more then two
observables per site was proposed and demonstrated for the case
of three particles and up to three observables per site. New tight
Bell inequalities for correlation functions and the respective
 conditions to satisfy then are found.
In particular, we present an inequality, (\ref{badziagineq}), in which all three observers perform three alternative measurements.
Interestingly, the inequalities (\ref{ineq1}), (\ref{ineq2}), (\ref{ineq3}) can be
shown to be implied by both the ones found in
\cite{wuzong,PLZ} and (\ref{badziagineq}). 
 
The method presented here can be generalized to $M>3$ settings Bell inequalities, and more than three parties. This definitely would lead to narrowing the class of states which allow a local realistic model. 
An old result of one of us \cite{ZUK93}, which shows a drastic constraint on such a class if the number of settings goes to infinity (for 4 parties or more), clearly supports this conjecture.

\section{Acknowledgments}
The work is part of EU 6FP programmes QAP and SCALA.
M. \.Zukowski was supported by Wenner-Gren Foundations.   M.
Wie\'sniak is supported by FNP stipends (START programme and within Professorial Subsidy 14/2003 for MZ). In the early stage of this work was MW was supported by a UG grant BW 5400-5-0260-4 and MZ by MNiI grant
1 P03B 04927.


\begin{thebibliography}{}
\bibitem{endnote} Please note, that $\Delta_0$ itself generates an inequality of the
CHSH-type.
\end{thebibliography}
\end{document}